\documentclass[aps,prd,twocolumn,groupedaddress,showpacs,showkeys]{revtex4}

\usepackage{graphicx}
\usepackage{subfigure}
\usepackage{epsfig}
\usepackage{amsfonts}
\usepackage{latexsym}

\begin{document}

%\preprint{UTTG-09-02}
\title{ UV-IR mixing and the quantum consistency of noncommutative gauge theories}

\author{Eric~Nicholson}
\email{ean@physics.utexas.edu}
\affiliation{Theory Group, Department of Physics\\
             University of Texas at Austin\\
             Austin, TX 78712, USA}

\date{\today}

\begin{abstract}
We study the quantum mechanical consistency of noncommutative
gauge theories by perturbatively analyzing the Wilsonian quantum
effective action in the matrix formulation. In the process of
integrating out UV states, we find new divergences having dual
UV-IR interpretations and no analogues in ordinary quantum field
theories. The appearance of these new UV-IR divergences has
profound consequences for the renormalizability of the theory. In
particular, renormalizability fails in any nonsupersymmetric
noncommutative gauge theory. In fact, we argue that
renormalizability generally fails in any noncommutative theory
that allows quantum corrections beyond one-loop.  Thus, it seems that noncommutative quantum theories are extremely sensitive to the UV, and only the
softest UV behavior can be tolerated.
\end{abstract}

\pacs{00} 

\maketitle

\section{Introduction}

By now, it has been a few years since it was realized that
noncommutative gauge theories emerge from string theory through
various decoupling limits \cite{sw:stri,sb:magn}. Nonetheless, the
understanding of the dynamics of noncommutative gauge theories
remains in a relatively primitive state compared to ordinary gauge
theories. The main difficulty is that noncommutative quantum field
theories are highly nonlocal. In fact, one can think of the
nonlocality as arising from elementary dipole degrees of freedom
whose transverse length is proportional to their center of mass
momentum \cite{sb:magn}. The novel behavior of these dipole quanta
gives rise to UV-IR mixing: UV dipoles grow long in spatial extent
and mediate instantaneous long distance interactions that dominate
in the IR.

Not surprisingly, UV-IR mixing leads to difficulties in the naive
application of conventional field theory techniques to
noncommutative theories. In particular, the non-decoupling of UV
and IR states leads to ambiguities in the distinction between
short and long distance physics. Consequently, there has been some
confusion regarding such things as the renormalization of UV
divergences \cite{ms:nonp,mr:uvdiv,cm:dual,cr:renorm}, the
treatment of IR divergences \cite{ms:nonp,mh:irdiv,rr:gaugefix},
and Wilsonian integration
\cite{vv:wilson,gp:wilsonrg,cw:statmech}. What is more, the
noncommutative gauge invariance is not preserved separately in
each diagram of the standard perturbative expansion; rather, gauge
invariance is achieved by an infinite resummation of diagrams
\cite{ki:inte,ar:uvir,hl:trek}. Finally, there is a sort of
naturalness problem with the conventional approach in that the
intrinsic dipole structure of the elementary field quanta is not
completely clear, although some suggestive results have been
obtained \cite{ki:inte}.

On the other hand, in the matrix formulation of noncommutative
gauge theory, the noncommutative gauge invariance is manifest
\cite{mv:mean}, as is the dipole character of the elementary
quanta \cite{jn:dipo}. In fact, as shown in \cite{jn:dipo}, the
matrix approach allows for a clear separation between the quantum
effects of UV and IR dipoles. In particular, it was shown that one
could make sense of Wilsonian integration despite UV-IR mixing.
For example, the interactions that result from integrating out the
UV states were explicitly calculated at both the one and two loop
order. The resulting interactions were found to dominate the long
distance behavior, which shed some light on UV-IR mixing in
noncommutative gauge theory, as well as the nonanalytic dependence
of the quantum theory on the noncommutativity parameter, $\theta$.
For a different point of view on how the matrix formulation
naturally leads to a bi-local representation, see
\cite{kk:bilocal}.

However, while some progress has been made, there are still many
unanswered questions. For one thing, the analysis of
\cite{jn:dipo} was limited to perturbation theory, which
generically breaks down due to strong quantum corrections.
Although it was conjectured in \cite{jn:dipo} that the
perturbative analysis of the Wilsonian quantum effective action is
valid for supersymmetric theories whose UV behavior is softer,
this remains to be proven. Furthermore, it is not even clear if
noncommutative quantum theories are renormalizable, although there
have been arguments for the affirmative \cite{cr:renorm}.

In this work, we continue to develop noncommutative gauge theory
in the matrix formulation. In section II, after reviewing the
basic ingredients of the dipole interpretation and the matrix
formulation of noncommutative gauge theory, we determine the
general form of the gauge invariant perturbative corrections to
the quantum effective action that result from integrating out UV
states. Based on this structure, we then show that the leading
long distance interactions cancel at each order in perturbation
theory for supersymmetric theories. In section III, we first
discuss renormalizability in the context of nonsupersymmetric
noncommutative theories and argue that, beyond one loop order,
UV-IR mixing generally introduces divergences that spoil the
consistency of the theory. We are led to conjecture that only
supersymmetric noncommutative theories that do not get
renormalized beyond one loop can be consistent quantum
mechanically. We end with some discussion of our results and some
concluding remarks. The appendices contain a review of the
important results of \cite{jn:dipo} and some other technical
details.

\section{Gauge invariant structure of perturbation theory}

Ultimately, we would like to further develop our intuition for how
UV-IR mixing affects the quantum mechanical consistency of
noncommutative gauge theories. To this end, it will prove
convenient to separate the degrees of freedom into an UV region,
defined by momenta much greater than any other  scale in the
problem, and an IR region defined by  all of the rest of momenta.
The strategy will be to first integrate out the UV modes in the
Wilsonian sense, and then to study the effective action in the IR
regime. In this section, we will discuss the general structure of
perturbative corrections to the effective action resulting from
integrating out UV states.

As discussed in \cite{hl:trek,ni:obse}, the structure of
perturbative corrections is strongly constrained by noncommutative
gauge invariance, which is a much larger symmetry group than
ordinary gauge symmetries. In fact, gauge invariant quantities
receive contributions from an infinite number of diagrams in terms of the
conventional perturbative expansion. For this reason, it is most
convenient to work in the matrix formulation of noncommutative
gauge theory, which is manifestly gauge invariant. The technical
machinery useful for performing perturbative calculations in the
matrix approach was developed in \cite{jn:dipo}, although the main
technical results from the matrix formulation, which will be used
extensively in what is to follow, are briefly reviewed in Appendix
\ref{ap:review}.

We will begin by recalling the interpretation of the matrix
degrees of freedom in terms of dipoles, since this viewpoint is
robust and will clearly generalize to all orders of perturbation
theory. The dipole interpretation follows naturally from the
representation of the matrix degrees of freedom in terms of the
noncommutative gauge group.  As discussed in \cite{ns:back}, the
degrees of freedom of noncommutative gauge theory are represented
as infinite dimensional  Hermitian matrices
\begin{equation}\label{eq:xmatrix}
X^{i}(t)=\hat{x}^{i}\otimes {\rlap{1} \hskip 1.6pt
\hbox{1}}_{N\times N}+ {\theta}^{ij}A_j(\hat{x},t),
\end{equation}
where $\hat{x}^{i}$ are time-independent Hermitian matrices
satisfying the algebra of the noncommuting $2p$-plane
\begin{equation}
[\hat{x}^{i},\hat{x}^{j}]=i\theta^{ij}{\rlap{1} \hskip 1.6pt
\hbox{1}}.
\end{equation}
The essential point is that $X^{i}$ lives in the adjoint
representation of the group noncommutative $U(N)$ which, as
indicated by the tensor product structure of
Eq.~(\ref{eq:xmatrix}), can be thought of as the product of
ordinary groups $U(\infty)\otimes U(N)$. The representations of
the ordinary $U(N)$ group will play the role of Chan-Paton factors
joined to the ends of the dipoles, and their treatment will come
naturally in our framework. The crucial observation is that the
adjoint representation of $U(\infty)$, furnished by functions of
$\hat{x}^{i}$, should be thought of in `t~Hooft double line
notation. In this language, the Feynman diagrams that appear in
the matrix formulation have a simple and intuitive interpretation
in terms of dipole degrees of freedom: \emph{the double lines
literally represent the spatial trajectory that is traced out by
the endpoints  of the dipole quanta as they propagate around the
loops} \cite{jn:dipo}.

In order to demonstrate and reinforce this dipole picture, we
shall review the results of explicit calculations and their
interpretations \cite{jn:dipo}. For example, consider the one loop
diagram
\begin{center}
\includegraphics[scale=.25]{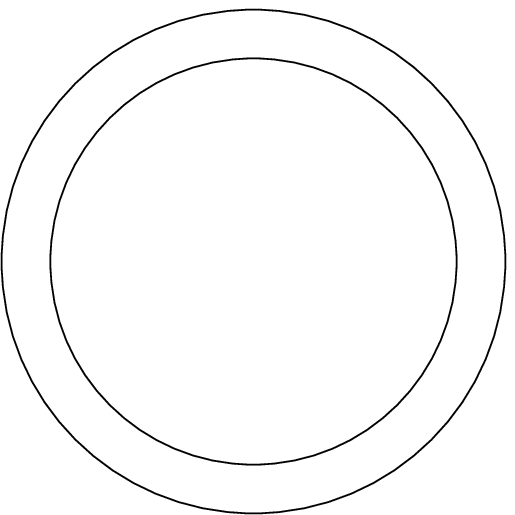}.
\end{center}
A simple matrix calculation gives an interaction term in the
effective Lagrangian proportional to
\begin{eqnarray}\label{eq:trace}
& & (N_B-N_F)\int d^{2p}x_1 d^{2p}x_2
\rho(x_1,t)\rho(x_2,t)\nonumber\\
& &
 \qquad\qquad\qquad\times\int\frac{d\omega}{2\pi}\log{\widetilde{G}(\omega,{\theta}^{-1}x_{12})},
\end{eqnarray}
where $N_B$ and $N_F$ are the numbers of bosonic and fermionic
polarization states, respectively, and the rest of the notation is
reviewed in Appendix \ref{ap:review}. Note that although we have
discussed only gauge field degrees of freedom explicitly, fermions
and other matter fields will generally contribute quantities of
the same form as the pure gauge field calculation, but they will
differ in the constant of proportionality as in
Eq.~(\ref{eq:trace}) above.

\begin{figure}\label{fig:cylinder}
\includegraphics[height=1cm]{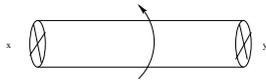}
\caption{High momentum virtual dipoles grow long in the transverse
direction and mediate instantaneous interactions between distant
background fluctuations at $x_1$ and $x_2$.} \label{fig:cyli}
\end{figure}

The $\log{\widetilde{G}(\omega,\theta^{-1}x_{12})}$ factor that
appears in the integrand of Eq.~(\ref{eq:trace}) is  familiar from
field theory if we identify the momentum of the virtual quantum
with $p_ i=\theta^{-1}_{ij}(x_1^{j}-x_2^{j})$. In fact, this is
precisely the relation between the center of mass momentum and the
end points of the dipoles that is expected \cite{sb:magn}! The
most obvious consequence of this relation is that UV dipoles are
long in spatial extent. Therefore, if we are to associate the
double lines of the matrix diagrams with the ends of the dipoles,
we should think of the one loop diagram above ``stretched out"
into a long thin cylinder as depicted in FIG.~\ref{fig:cyli}.

The quantity $\rho(x,t)$ which appears in Eq.~(\ref{eq:trace}) is
a gauge invariant functional of the low momentum background field
that is well localized in space around $x$, given the fact that we
are only integrating out high momentum states, as discussed in
\cite{jn:dipo}. Evidently, we are to associate a trace, which
produces the gauge invariant functional $\rho$, to each boundary
of the cylinder. In fact, this leads to the connection between
Eq.~(\ref{eq:trace}) and the field theory diagrams that appear in
the conventional perturbative expansion. If we expand
$\rho(x)={\rm tr}_{N}({\rlap{1} \hskip 1.6pt
\hbox{1}})+\Delta(x)$, then as discussed in \cite{jn:dipo}, we
identify the constant term as the contribution from field theory
diagrams with no background field insertions on the corresponding
boundary, while the term with nontrivial position dependence
descends from diagrams that include background gauge field
insertions on the corresponding boundary. This interpretation is
supported by the fact that $\Delta(x)=0$ for field configurations
gauge equivalent to $A_i(x)=0$. Using this reasoning, we can
conclude that the vacuum diagram, which involves no insertions on
either boundary, is proportional to
\begin{equation}\label{eq:vacuum}
(N_B-N_F)N^{2}V\int\frac{d\omega
d^{2p}p}{2\pi(2\pi)^{2p}}\log\widetilde{G}(\omega,p).
\end{equation}
Of course, this is just the expected result from field theory.

The contribution from diagrams with planar background field
insertions vanishes because $\int d^{2p}x\Delta(x,t)=0$; however,
the nonplanar diagrams yield a nontrivial interaction term
proportional to
\begin{eqnarray}\label{eq:delta}
& & (N_B-N_F)\int d^{2p}x_1 d^{2p}x_2
\Delta(x_1,t)\Delta(x_2,t)\nonumber\\
& &
\qquad\qquad\qquad\times\int\frac{d\omega}{2\pi}\log{\widetilde{G}(\omega,{\theta}^{-1}x_{12})}.
\end{eqnarray}
Apparently, Eq.~(\ref{eq:delta}) describes the instantaneous
interaction between distant background fluctuations at $x_1$ and
$x_2$, as illustrated in FIG.~\ref{fig:cyli}. In fact, this
interaction term demonstrates the novel feature of noncommutative
quantum field theories: UV dipoles grow long in spatial extent and
mediate instantaneous long distance interactions, which is how we
can think about UV-IR mixing.

However, it should be noted that, in the case
of Eq.~(\ref{eq:delta}), the interactions grow strong at long
distances because
\begin{equation}\label{eq:strong}
\int
\frac{d\omega}{2\pi}\log\widetilde{G}(\omega,{\theta}^{-1}x_{12})\sim|x_1-x_2|+\textrm{const}.
\end{equation}
As discussed in \cite{mv:mean}, the strong interaction expressed
in Eq.~(\ref{eq:strong}) is due to the leading IR poles in
external momenta that result from the UV region of loop integrals
involving nonplanar background insertions. The appearance of
nonanalytic behavior in external momenta has also been discussed
from the field theory perspective in \cite{ls:gaug}. In the
particular case of Eq.~(\ref{eq:strong}), we will require that
$N_B=N_F$ so that this term vanishes and perturbation theory
remains valid. However, the appearance of strong IR corrections
from the quantum effects of virtual UV states and their treatment
in general will be a central theme of this work.

Furthermore, as discussed in \cite{jn:dipo}, it is important to
realize that the one loop matrix calculation (\ref{eq:trace})
gives only the leading term in a derivative expansion involving
the background field. The next to leading order one loop diagram
involves two extra insertions of the background field strength
\begin{center}
\includegraphics[scale=0.25]{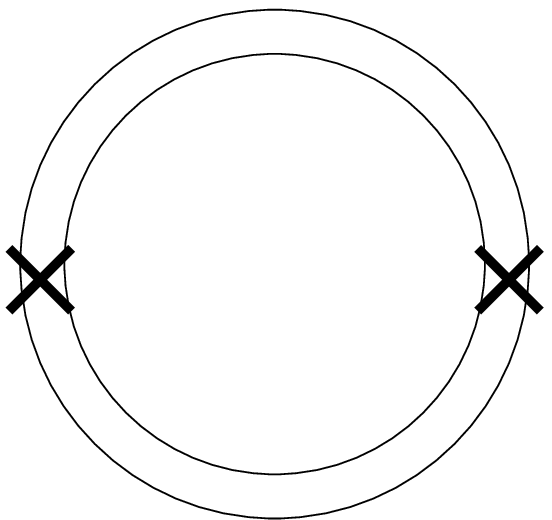}.
\end{center}
The contribution from this diagram was shown to be of the form
\begin{eqnarray}\label{eq:ntlo}
& & \int d^{2p}x_1
d^{2p}x_2\left[c_1{\rho}_{FF}(x_1,t)\rho(x_2,t)+c_2{\rho}_{F}(x_1,t)\right.\\
& &
\qquad\times\left.{}{\rho}_{F}(x_2,t)\right]\int\frac{d\omega}{2\pi}
\widetilde{G}({\omega},{\theta}^{-1}x_{12})\widetilde{G}({\omega},{\theta}^{-1}x_{21})\nonumber,
\end{eqnarray}
where $c_1$ and $c_2$ are constants depending on the
dimensionality and the matter content of the theory, as previously
discussed. Notice that a nontrivial contribution from planar field theory diagrams is contained in the first term above.

We will discuss Eq.~(\ref{eq:ntlo}) in much more detail in the next section, but for now, we shall just remark on its basic features. Evidently, the field strength operators that were inserted into
the matrix diagram above appear as operators attached to the end
of Wilson lines. This structure seems to reinforce the
correspondence between the boundaries of the matrix diagram and
traces which lead to gauge invariant Wilson lines. In general,
higher order terms in the derivative expansion involve higher
dimensional operator insertions into the loop diagram, and hence,
the Wilson lines. However, the double trace structure and the
corresponding  interpretation in terms of long distance two-body
interactions remains the same. Moreover, it is interesting  that the appearance of the
propagator $\widetilde{G}$ in Eq.~(\ref{eq:ntlo}) is again
consistent with our field theory intuition.

\begin{figure}
 \subfigure[An illustration of the contribution
from the first order treatment of quartic interaction terms given
by
(\ref{eq:threebody}).]{\label{quartic:subfig:a}\begin{minipage}[b]{.5\textwidth}\centering
\includegraphics[scale=0.5]{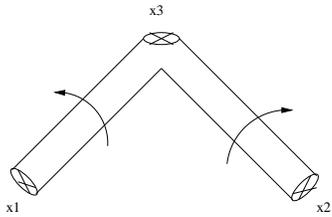}\end{minipage}}

\subfigure[An illustration of the contribution  from the second order treatment of cubic
interaction terms given by
(\ref{eq:cubicthreebody}).]{\label{quartic:subfig:b}\begin{minipage}[b]{.5\textwidth}\centering
\includegraphics[scale=0.5]{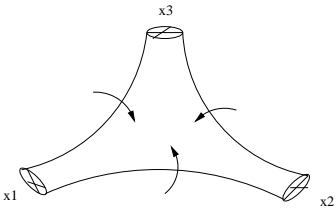}\end{minipage}}
\caption{Long distance three-body interactions corresponding to
high momentum dipoles propagating in two-loop
diagrams.}\label{fig:twolooppicture}
\end{figure}

Higher loop diagrams, on the other hand, will involve more traces,
and will therefore lead to multi-body interactions. For example
the leading two-loop contribution to the effective Lagrangian
comes from
\begin{center}
\begin{minipage}[c]{2.7cm}
\includegraphics[scale=0.4]{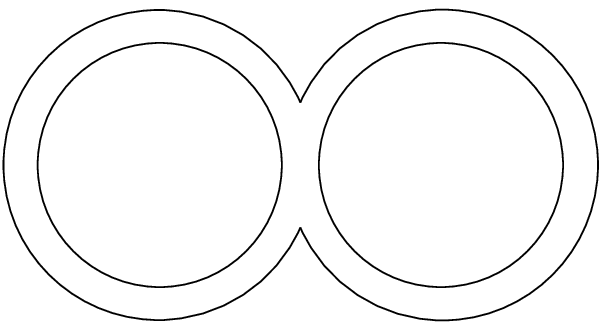}
\end{minipage}%
\begin{minipage}[c]{0.5cm}
$+$
\end{minipage}%
\begin{minipage}[c]{2.7cm}
\includegraphics[scale=0.3]{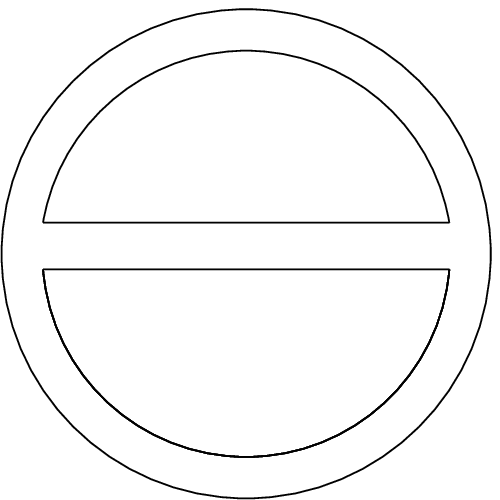}.
\end{minipage}
\end{center}
 Note that we have not included the
nonplanar matrix diagrams because their contribution to the
Wilsonian integration is exponentially suppressed due to Moyal
phase factors \cite{jn:dipo}. As indicated by the double line diagrams above, the leading
two-loop diagrams give triple trace contributions. The first
diagram gives an interaction term in the effective Lagrangian
proportional to \cite{jn:dipo} {\setlength\arraycolsep{2pt}
\begin{eqnarray}\label{eq:threebody}
 & & \int d^{2p}x_1 d^{2p}x_2 d^{2p}x_3 \rho(x_1,t)\rho(x_2,t)\rho(x_3,t)\int
 \frac{d{\omega}_1}{2\pi}\frac{d{\omega}_2}{2\pi} \nonumber\\
 & & \qquad\times \widetilde{G}({\omega}_1,{\theta}^{-1}x_{13})\widetilde{G}({\omega}_2,{\theta}^{-1}x_{23})
 \end{eqnarray}}\\
and, as calculated in Appendix \ref{ap:twoloop}, the second gives
a term  proportional to
\begin{eqnarray}\label{eq:cubicthreebody}
& & \int d^{2p}x_1 d^{2p}x_2
d^{2p}x_3\rho(x_1,t)\rho(x_2,t)\rho(x_3,t)
\int\frac{d\omega_1}{2\pi}\frac{d\omega_2}{2\pi}\nonumber\\
& & \qquad\times(\omega_1\omega_2-x_{12}\cdot
x_{23})\widetilde{G}(\omega_1,\theta^{-1}x_{12})\nonumber\\
& & \qquad\times\widetilde{G}(\omega_2,\theta^{-1}x_{23})
\widetilde{G}(\omega_1+\omega_2,\theta^{-1}x_{31}).
\end{eqnarray}
Indeed, as expected based on our intuition, Eqs.~(\ref{eq:threebody}) and (\ref{eq:cubicthreebody}) describe long distance interactions that arise from high momentum dipoles growing large in spatial extent and ``stretching out'' the matrix diagrams, as depicted in FIG.~\ref{fig:twolooppicture}. Furthermore, these expressions bear a close resemblance to what is
expected from ordinary field theory, and in fact, the general
structure of perturbative corrections is starting to emerge.

To each boundary of the double line diagram we should associate a
point in space and a trace which yields a gauge invariant Wilson
line. The position dependence of the interaction strength between
the Wilson lines follows from an integral over frequencies with
the integrand being given by a particular function of both the
dipole frequencies and momenta as defined by the separation between
the space points associated to the boundaries of the double line
diagram. In fact, the particular function of dipole frequencies
and momenta corresponds precisely to the structure of momentum
space field theory propagators that appear in the analogous
process in ordinary field theory.

The subleading terms in the derivative expansion will have a
similar trace structure but Wilson lines modified to include
higher derivative operator insertions as in Eq.~(\ref{eq:ntlo}).
Naturally, the subleading terms will include a different function
of dipole frequencies and momenta that reflects the extra
propagators that are required by the operator insertions. Note
that for dipole degrees of freedom, powers of momentum are
equivalent to powers of separation, as discussed in \cite{jn:dipo}
and in Appendix \ref{ap:review}. Therefore, the inclusion of more
powers of momentum in the denominator leads to faster falloff with
distance, which is expected from subleading terms in a derivative
expansion. However, quantum corrections coming from higher loop
orders are generally strong, as we will discuss in the next
section.

Of more importance to our immediate goal, is the
determination of the form of the leading long distance
interactions at each order in perturbation theory. It is clear
from the general structure discussed above that at $L$-loop order
the leading long distance interactions will involve $L+1$ Wilson
lines with no operator insertions, such as Eq.~(\ref{eq:trace}) in
the case $L=1$ and both Eq.~(\ref{eq:threebody}) and
Eq.~(\ref{eq:cubicthreebody}) in the case $L=2$. The crucial
observation is that since the vacuum diagrams are contained in
interactions of this form, as shown explicitly by
Eq.~(\ref{eq:vacuum}) in the one-loop case, these terms must
vanish entirely if the vacuum diagrams vanish. Thus, \emph{in
supersymmetric noncommutative theories, the leading long distance interactions
must cancel at each order in perturbation theory}.

This statement is a generalization to all loop orders of the
cancellation that we have discussed in the case of the leading
one-loop interactions given by Eq.~(\ref{eq:trace}). It reflects
the fact that supersymmetric theories are softer in the UV, and
hence, the IR behavior that is generated by UV-IR mixing is not as
strong. In fact, the cancellation of the leading IR interactions
occurs even in minimally supersymmetric theories. The relation
between the amount of supersymmetry and the cancellation of other
terms has been discussed in
\cite{ls:gaug,vv:wilson,dz:susy,rr:gaugefix} from the field theory
perspective, and we will touch on this important point as well in
the next section.

In summary, we have determined the gauge invariant form of
perturbative corrections to the Wilsonian quantum effective action
for noncommutative gauge theories. It followed immediately from
the structure of perturbation theory that the leading long
distance interactions cancel order by order in the supersymmetric
theories. Since nonsupersymmetric theories do not enjoy these
cancellations, they are strongly coupled at long distances. In
fact, what we are beginning to discover is that noncommutative
quantum theories are highly sensitive to the UV, which we will
further explore in the next section.

\section{Renormalizability}

In the previous section, we determined the gauge invariant form of
the perturbative quantum corrections to the effective action that
result from integrating out UV states. In this section we will
take a closer look at these perturbative corrections and their
divergence properties. In addition to the expected types of UV
divergences, we will find new divergences that have dual UV and IR
interpretations and for which there are no analogues in ordinary
quantum field theories. In fact, we will argue that these new
UV-IR divergences, if they are allowed, will spoil the
renormalizability of the theory. In the end, we will find that
only supersymmetric theories with the softest UV behavior can be
consistent quantum mechanically.

Because it will set the ground work for the discussion of the more
interesting supersymmetric case, we will start this section by
discussing  divergences in the context of nonsupersymmetric
noncommutative gauge theories. We will begin with the familiar
procedure of gauge coupling renormalization. In fact, in this
case, it will turn out that renormalizability is threatened
already at the two loop level. To see how this happens, we first
recall from \cite{jn:dipo} the form of the one-loop quantum
correction to the gauge coupling. The only candidate for this type
of term is Eq.~(\ref{eq:ntlo}) because it comes from a one-loop
graph with two insertions of the field strength operator. To
isolate the UV divergence in the planar diagrams, we must use the
splitting scheme $\rho(x)={\rm tr}_{N}({\rlap{1} \hskip 1.6pt
\hbox{1}})+\Delta(x)$ and focus on the first term, as previously
discussed. The integral then factorizes into
\begin{eqnarray}
 & & c_1\int d^{2p}x_1{\rho}_{FF}(x_1,t)N\int\frac{d{\omega}}{2\pi}d^{2p}x_{12}\widetilde{G}({\omega},{\theta}^{-1}x_{12})^{2}\nonumber\\
 &=&  c_1 {\rm Tr}\left({[B^{i},B^{j}]}^{2}\right) N\int\frac{d{\omega}d^{2p}p}{2\pi(2\pi)^{2p}}{\widetilde{G}(\omega,p)}^{2}.
\end{eqnarray}
This quantity is easily recognized as contributing to the
renormalization of the operator ${\rm Tr}[B^{i},B^{j}]^{2}$, as
expected. Therefore, the one-loop quantum correction to the gauge
coupling is proportional to the dimensionless quantity
\begin{equation}\label{eq:gaugerenorm}
 g^{2}N\int\frac{d{\omega}d^{2p}p}{2\pi(2\pi)^{2p}}{\widetilde{G}(\omega,p)}^{2},
\end{equation}
where we have restored the gauge coupling $g^{2}$ explicitly.

If we are to consistently absorb the one-loop correction
(\ref{eq:gaugerenorm}) into a new quantum corrected gauge
coupling, we must find the same quantity at higher loops
correcting each factor of $g^{2}$ that appears. Otherwise,
renormalizability will be threatened, because we will be forced to
introduce new couplings into the effective Lagrangian in order to
cancel all of the resulting divergences. However, in our
conventions, the dependence on the gauge coupling is particularly
simple -- powers of $g^{2}$ count the number of loops. Thus, powers
of the gauge coupling only appear multiplying the loop
corrections; the Wilson lines, which encode the background gauge
field insertions into the loops, are independent of the coupling.
Therefore, to test renormalizability of the gauge coupling at the
two loop order, we should look for a term of the form
\begin{eqnarray}\label{eq:required}
& & \int d^{2p}x_1 d^{2p}x_2
\rho(x_1,t)\rho(x_2,t)|x_1-x_2|\nonumber\\
& &
\qquad\times\int\frac{d{\omega}d^{2p}p}{2\pi(2\pi)^{2p}}{\widetilde{G}(\omega,p)}^{2},
\end{eqnarray}
because it corresponds to the one-loop correction to $g^{2}$ given
by Eq.~(\ref{eq:gaugerenorm}) multiplying the leading one-loop
interaction (\ref{eq:trace}).

In order to find this sort of term, we consider the leading
two-loop interactions. As discussed in Appendix \ref{ap:twoloop},
these contributions can be combined into the form of
Eq.~(\ref{eq:threebody}). It is now straight forward to isolate
the divergent quantum corrections to the two-body interactions.
The constant term in either $\rho(x_1)$ or $\rho(x_2)$ gives
\begin{eqnarray}\label{eq:propcorrection}
 & & \int d^{2p}x_1 d^{2p}x_3 \rho(x_1,t)\rho(x_3,t)\int \frac{d\omega_1}{2\pi}\widetilde{G}(\omega_1,{\theta}^{-1}x_{13})\nonumber\\
 & & \quad\times\int\frac{d\omega_2 d^{2p}p_2}{2\pi(2\pi)^{2p}}\widetilde{G}(\omega_2,p_2).
\end{eqnarray}
Apparently, Eq.~(\ref{eq:propcorrection}) involves the leading
one-loop UV divergent contribution to the mass term of the field theory propagator
$\widetilde{G}$ appearing in Eq.~(\ref{eq:trace}). In fact, this
type of correction is familiar from ordinary quantum field
theories. However, it will turn out that in the supersymmetric theories of ultimate interest to us, corrections of this form will not appear. In any case,
Eq.~(\ref{eq:propcorrection}) is not the term we are looking for,
so we move on.

The only other candidate for the appearance of
Eq.~(\ref{eq:required}) in the leading order two-loop diagrams is
the interaction that comes from the constant term of $\rho(x_3)$
\begin{eqnarray}\label{eq:twoloopcorrection}
 & & \int d^{2p}x_1 d^{2p}x_2 \rho(x_1,t)\rho(x_2,t)\int\frac{d\omega_1}{2\pi}\\
& & \quad\times\int\frac{d\omega_2
d^{2p}p_3}{2\pi(2\pi)^{2p}}\widetilde{G}(\omega_{21},p_3-{\theta}^{-1}x_{12})\widetilde{G}(\omega_2,p_3)\nonumber.
\end{eqnarray}
Needless to say, Eq.~(\ref{eq:twoloopcorrection}) is not the term
we need either; thus, the renormalizability of the gauge coupling
is threatened already at the two-loop order. However, a closer
look at Eq.~(\ref{eq:twoloopcorrection}) reveals an entirely new
problem facing the renormalizability of noncommutative theories
that we will now discuss.

Evidently, Eq.~(\ref{eq:twoloopcorrection}) represents a two-loop
quantum correction to the leading one-loop two-body interaction
(\ref{eq:trace}). The growth in $|x_1-x_2|$ of the interaction
strength of Eq.~(\ref{eq:twoloopcorrection}) is given by
\begin{eqnarray}\label{eq:intcorrection}
\int\frac{d\omega_1}{2\pi}\frac{d\omega_2
d^{2p}p_3}{2\pi(2\pi)^{2p}}\widetilde{G}(\omega_{21},p_3-{\theta}^{-1}x_{12})\widetilde{G}(\omega_2,p_3).
\end{eqnarray}
Clearly, Eq.~(\ref{eq:intcorrection}) is UV divergent for $p\geq
1$, so we must introduce an UV cutoff $M$. The integral can now be
calculated in a straight forward fashion and the leading
dependence on separation goes as
$|x_1-x_2|^{2p-2}\log(|x_1-x_2|^{2}/{M^{2}})$. However, since
there is no two-body interaction that grows as $|x_1-x_2|^{2p-2}$
at one loop, this logarithmic divergence cannot be absorbed into
the gauge coupling. Therefore, a new coupling for a two-body
interaction of the form
\begin{equation}
\int d^{2p}x_1d^{2p}x_2\rho(x_1)\rho(x_2)|x_1-x_2|^{2p-2}
\end{equation}
must be introduced into the effective Lagrangian so that the
divergence in Eq.~(\ref{eq:twoloopcorrection}) can get cancelled.

Although one can arrange for the cancellation of divergent
corrections to any interaction strength by continuing to introduce
new interactions with divergent couplings, as above, it is not
hard to see that this procedure is a serious threat to the
renormalizability of the theory. To see this, we need only note
that, at higher orders, there will be quantum corrections to the
interaction strengths that come in the form of powers of the
dimensionless quantity
\begin{equation}\label{eq:running}
\frac{g^{2}}{|x-x^{\prime}|^{3-2p}}\log(|x-x^{\prime}|^{2}/{M^{2}}).
\end{equation}
Therefore, to cancel all of the divergences that appear in the
various powers of (\ref{eq:running}), we would have to introduce
an infinite number of new interactions. The reason is that there
are an infinite number of distinct functions of $|x-x^{\prime}|$
that can appear, each requiring a different interaction term with
its own coupling in order to cancel the required divergences in
the effective Lagrangian. Thus, it seems that divergent
corrections to the interaction strength between Wilson lines, a
new feature of noncommutative theories having no analogue in
ordinary quantum field theories, will generally spoil the
renormalizability of the theory.

A crucial assumption in the argument above is that if divergent
quantum corrections to interaction strengths occur at the two-loop
order, they will appear at all higher orders as well. In fact,
this is easy to see graphically by recalling the relation between
matrix diagrams and field theory diagrams. Using the
correspondence that we discussed earlier in section II, we find
that Eq.~(\ref{eq:twoloopcorrection}) descends from the leading
order terms in the external momenta expansion of field theory
diagrams of the form
\begin{center}
\begin{minipage}[c]{2.7cm}
\includegraphics[scale=0.4]{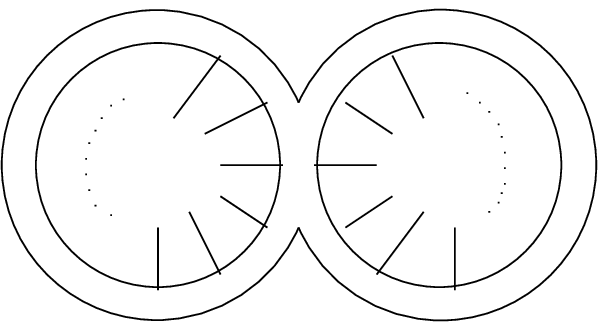}
\end{minipage}%
\begin{minipage}[c]{0.5cm}
\end{minipage}%
and
\begin{minipage}[c]{2.7cm}
\includegraphics[scale=0.4]{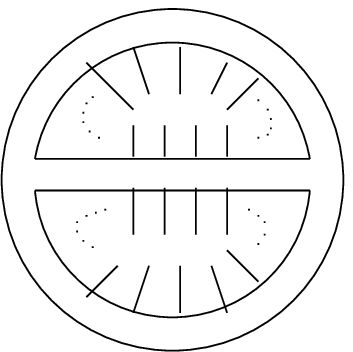}.
\end{minipage}
\end{center}

\begin{figure}
\subfigure[Some three-loop field theory diagrams containing
dangerous quantum corrections to the two-loop three-body
interactions.]{\label{three:subfig:a}\begin{minipage}[b]{.5\textwidth}\centering
\includegraphics[scale=0.5]{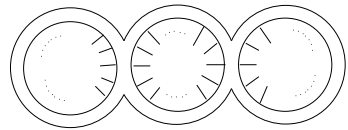}\end{minipage}}

\subfigure[Some three-loop field theory diagrams containing
dangerous quantum corrections to the one-loop two-body
interactions.]{\label{three:subfig:b}\begin{minipage}[b]{.5\textwidth}\centering
\includegraphics[scale=0.5]{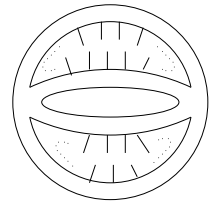}\end{minipage}}
\caption{Examples of three-loop nonplanar field theory diagrams
that contribute dangerous quantum corrections to the strength of
lower order interactions.}\label{fig:threeloop}
\end{figure}

Evidently, the dangerous divergences are not due to planar
subdiagrams, which have the standard UV interpretation, but are
due to purely nonplanar field theory diagrams. We will develop
more understanding for what this means in the moment, but for now
let us note that higher order graphs of this type exist at each
loop order in perturbation theory. For example,
FIG.~\ref{fig:threeloop} depicts some three-loop field theory
diagrams that give rise to divergent quantum corrections to
three-body and two-body interactions. Thus, dangerous corrections
to interaction strengths that first appear at some given order in
perturbation theory will receive quantum corrections from all
higher loop orders, generating corrections in the form of
powers of (\ref{eq:running}).

However, by the arguments above, the divergences appearing in
powers of Eq.~(\ref{eq:running}) will inevitably spoil
renormalizability. Therefore, any renormalizable noncommutative
theory cannot allow the leading order terms in the derivative
expansion, which as we have seen, contain these dangerous
corrections. We are forced to conclude that \emph{only
supersymmetric noncommutative theories can be renormalizable}
since, in this case, the leading order interactions always vanish.

Before discussing the supersymmetric theories, let us take some
time to understand the nature of the dangerous divergences by
recasting them into the form of momentum integrals. For example,
consider the transformation of Eq.~(\ref{eq:twoloopcorrection}) to
momentum space
\begin{eqnarray}\label{eq:moyal}
& & \int
\frac{d^{2p}k}{(2\pi)^{2p}}\widetilde{\rho}(k,t)\widetilde{\rho}(-k,t)\frac{d\omega_1d^{2p}q}{2\pi(2\pi)^{2p}}\frac{d\omega_2d^{2p}p}{2\pi(2\pi)^{2p}}\nonumber\\
& & \qquad\times e^{ik\cdot\theta\cdot
q}\widetilde{G}(\omega_{21},p-q)\widetilde{G}(\omega_2,p).
\end{eqnarray}
We can first perform the integrals over $\omega_2$ and $p$ by
introducing a Schwinger parameter and an UV cutoff $M$, as in
\cite{ms:nonp}. Then the integral over $\omega_1$ and $q$ is
regular, and we are left with a quantity proportional to
\begin{eqnarray}
& & \int
\frac{d^{2p}k}{(2\pi)^{2p}}\frac{\widetilde{\rho}(k,t)\widetilde{\rho}(-k,t)}{(
k^{2}+1/{M^{2}})^{2p-1}}\nonumber\\
&=& \int d^{2p}x_1d^{2p}x_2 \rho(x_1,t)\rho(x_2,t)\nonumber\\
& & \qquad\times\int
\frac{d^{2p}k}{(2\pi)^{2p}}\frac{e^{ik\cdot(x_1-x_2)}}{(
k^{2}+1/{M^{2}})^{2p-1}}.
\end{eqnarray}
Now the $M\rightarrow\infty$ limit results in an IR divergence
instead of an UV divergence, and $1/M$ plays the role of an IR
cutoff! The transform back into position space just gives the same
correction to the interaction strength that we have already
determined directly from Eq.~(\ref{eq:intcorrection}).

Apparently, these types of divergences have a dual UV-IR
interpretation. The origin of this duality can be traced to the
Moyal phase factor $\exp(ik\cdot\theta\cdot q)$ in
Eq.~(\ref{eq:moyal}), which represents the nonplanarity of the
background insertions and gives rise to UV-IR mixing. In
particular, the UV loop integrals over virtual states, labelled by
momenta $p$ and $q$, produce nonanalytic dependence on the
background momentum $k$ that becomes  important in the IR. Thus, on one
hand, the divergence comes from integrating over UV states, but on
the other hand, the divergence cast be recast into the form of an
IR singular Fourier transform.

Actually, as discussed in \cite{jn:dipo}, this type of UV-IR
mixing has a simple interpretation in terms of dipole degrees of
freedom: the separation between the endpoints of the dipoles is the dual
variable to the external momenta $k$; therefore, the IR singular
Fourier transform gives rise to corrections to interaction
strengths which grow strong with large separation; but, since
powers of separation are equivalent to powers of dipole momentum,
these strong corrections can also be thought of as arising from UV
divergences in the theory. Thus, the dipole intuition that emerges from the matrix formulation seems to shed new light on the proposal of
\cite{cm:dual,dz:susy} concerning the dual interpretation of
divergences due to nonplanar diagrams.

In fact, at this point, let us take an aside to remark on the virtues of
the matrix formulation of noncommutative gauge theory.  First of
all, the matrix degrees of freedom naturally encode the elementary
dipole structure of the field quanta, and hence, the essence of
UV-IR mixing. This is a very convenient aspect of the matrix
approach, which leads to great conceptual clarity in the physics
of noncommutative quantum theories, as we have seen. Perhaps, the
most important advantage of the matrix description, however, is the manifest
nature of the noncommutative gauge invariance. This crucial
property allows us to distinguish between various long distance
interactions, which in turn, forms the basis for our
interpretation of the UV-IR divergences that appear in the theory.

Back to the issue of divergences in supersymmetric noncommutative
theories, we now consider interactions that are of next to leading
order in the derivative expansion. As we have discussed, these are
the lowest order terms that do not generally vanish in the case of
supersymmetric theories. However, we will find that UV-IR
divergences do generally occur in the next to leading order
interactions, and therefore, supersymmetric theories
generally suffer from UV-IR divergences, as well. For example, as derived in Appendix~\ref{ap:ntlotl}, the next to
leading order two-loop interactions include dangerous terms such
as Eq.~(\ref{eq:dangerous}), which contain a quantum correction
to the interaction strength of Eq.~(\ref{eq:ntlo}) that goes as
\begin{eqnarray}\label{eq:div}
& & \int
d\omega_1d\omega_2d^{2p}x_3\widetilde{G}(\omega_1,\theta^{-1}x_{12})^{2}\widetilde{G}(\omega_2,\theta^{-1}x_{23})\nonumber\\
& &
\qquad\times\widetilde{G}(\omega_1+\omega_2,\theta^{-1}x_{13}).
\end{eqnarray}
Following an analysis parallel to the one above for
Eq.~(\ref{eq:intcorrection}), it is straight forward to show that
UV-IR divergences appear in Eq.~(\ref{eq:div}) for $p\geq 3/2$
(Note that by $p$ equal to a half-integer we really mean that, in
addition to $p-1/2$ noncommutative two-planes, there is an
additional commutative spatial direction. The proper way to
include this case in our analysis is to let $p$ remain the integer valued
number of noncommutative two-planes and replace $d\omega$ with
$d^{2}\omega$ to represent the two commuting spacetime
directions. All of the conclusions that we have drawn up to this point are true also in the case when the space dimensionality is odd).

Furthermore, the graphical argument that we employed earlier in
the case of the leading order interactions, can be easily
generalized to include the next to leading order terms in the
derivative expansion. We need only consider adding higher
derivative operator insertions in the matrix framework or, in the
field theory picture as in FIG.~\ref{fig:threeloop}, taking the
next to leading order terms in the expansion in external momenta.
In any case, we are forced to conclude that if dangerous UV-IR
terms appear at two loops, then there will be more dangerous
quantum corrections that emerge from all higher loop orders.
Therefore, if Eq.~(\ref{eq:dangerous}) appears in $3+1$ dimensions
or higher, UV-IR divergences will proliferate and renormalizability
will fail.

In order to determine what corrections are allowed by
supersymmetric theories, let us take some time to discuss the
proportionality constants that we have so far ignored. This will
be fruitful because we will find that the nonrenormalization
theorems enjoyed by ordinary supersymmetric theories have a
natural extension to noncommutative theories. We can then use
these nonrenormalization theorems to show that $3+1$ dimensional
noncommutative theories with a sufficiently high degree of
supersymmetry do not allow UV-IR divergences, and therefore,
renormalizability does not face any obvious obstruction in these
special theories.

We begin by considering the content of the matrix diagrams from a
field theory point of view. First of all, the double line matrix
diagrams themselves only contain information pertaining to the
contraction of gauge indices, which encodes the planarity and
nonplanarity of both the external operator insertions and the
internal propagators. Thus, the double line diagrams contain
purely topological information, not specific to any particular
gauge theory -- ordinary or noncommutative. In terms of the
framework that we have discussed up to this point, this
topological information only fixes the ratio of coefficients
between terms with a similar propagator structure but different
Wilson line structure. For example, ${c_1}/{c_2}$ is fixed by
topological considerations, as are both ${b_6}/{b_7}$ and
${b_6}/{b_8}$. Furthermore, the coefficients coming from nonplanar
matrix diagrams, which we have not discussed here, are fixed
relative to the coefficients of the corresponding planar diagrams
(Note that factors of $N$ are contained in the traces of the
Wilson lines, not the coefficients that we are referring to).
However, topological quantities, such as the ratio between
coefficients, are independent of any characteristic specific to
the theory in question such as the field content, the allowed
couplings between fields, and whether the theory is noncommutative
or not.

Although the ratios of certain coefficients are topological
quantities, independent of the theory in question, the
coefficients coming from planar and nonplanar matrix diagrams
involving a similar propagator structure all share a common
normalization factor that depends on the number and type of modes
that propagate around the loops. Thus, the overall normalization
factor does depend on characteristics specific to a particular
theory, such as the field content, the allowed couplings between
fields, and the coupling constants. However, the noncommutativity
of the theory enters only through the Moyal phase factors inside
the momentum integrals of nonplanar field theory diagrams;
noncommutativity does not affect the coefficient outside the
integral. It follows that a given coefficient in a particular
noncommutative gauge theory is identical to the same coefficient
in the corresponding ordinary gauge theory. After all, with the
exception of Moyal phase factors in the the noncommutative
theories, the contraction of fields is the same in both cases.
Therefore, \emph{the nonrenormalization theorems of ordinary
supersymmetric theories generalize immediately to noncommutative
supersymmetric theories}.

As an application of this useful result, let us return to the
problem of dangerous UV-IR divergent terms in the next to leading
order two-loop contribution to the quantum effective action, which
are given by Eq.~(\ref{eq:dangerous}). According to our reasoning
above, $b_6$, $b_7$, and $b_8$ are all proportional to the
two-loop contribution to the beta function. Therefore, we can
guarantee their vanishing by imposing $\mathcal{N}=2$
supersymmetry in $3+1$ dimensions. In this case, the dangerous
two-loop interactions cancel, as do all higher loop corrections.
Moreover, the two-loop terms given by Eq.~(\ref{eq:leading})
cancel because no mass terms are allowed to appear perturbatively,
given any amount of supersymmetry. Thus, renormalizability for
$3+1$ dimensional noncommutative gauge theories requires at least
$\mathcal{N}=2$ supersymmetry, in which case the  next to leading
order interactions are restricted to the one-loop result given by
Eq.~(\ref{eq:ntlo}), along with the two-loop terms containing the
one-loop correction to the gauge coupling (\ref{eq:beta}). In the
case of $\mathcal{N}=4$ supersymmetry, none of the next to leading
order interactions are allowed what so ever.

Moreover, it is also worth mentioning that $\mathcal{N}\geq 2$
supersymmetry also implies that most, if not all, of the next to
leading order nonplanar matrix diagrams cancel as well. The reason
is that, as we have discussed, the coefficients of these
contributions differ from their planar counterparts only by a
combinatoric factor. Thus, the only next to leading order
nonplanar remnant is
\begin{center}
\includegraphics[scale=.4]{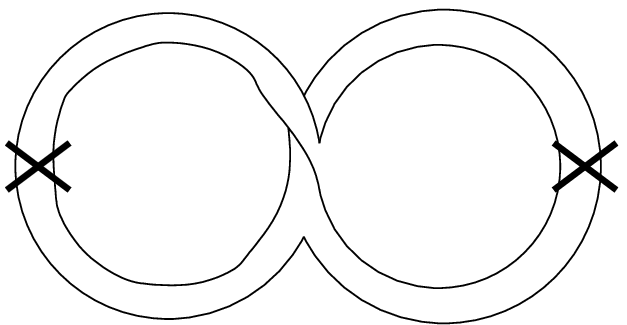}
\end{center}
which is proportional to the one-loop beta function. Of course,
all of the leading order nonplanar diagrams cancel, given any
amount of supersymmetry. The advantage of eliminating the leading
and next to leading order nonplanar matrix diagrams, is that they
are divergent when the Wilsonian cutoff is removed. In fact, these
divergences resulting from  the IR region of integration in
nonplanar loop integrals, have posed a major challenge to our
understanding of the quantum effects of IR states in
noncommutative theories. Therefore, it is fortunate that
renormalizability of UV quantum fluctuations turns out to also
constrain most of these  IR divergences to vanish as well.
Although a thorough treatment of the quantum effects of IR states
is beyond the scope of this work, it seems that our results here
are already a help in this direction.

While we have shown that $\mathcal{N}\geq 2$ supersymmetry is
sufficient to prevent UV-IR divergences from appearing in diagrams
containing two insertions of the field strength, we should still
worry about dangerous terms appearing from diagrams involving
other operator insertions. For example, all orders of wave
function renormalization for matter hypermultiplets is generally
allowed by $\mathcal{N}=2$ theories, although $\mathcal{N}=4$
theories do not allow any renormalization at all. Therefore,
$\mathcal{N}=2$ gauge theories generally involve new UV
divergences at each loop order, which will inevitably lead to the
proliferation of dangerous UV-IR divergences via UV-IR mixing.
After all, the divergent integral
\begin{equation}
\int
\frac{d\omega_2d^{2p}p_3}{2\pi(2\pi)^{2p}}\widetilde{G}(\omega_2,p_3)\widetilde{G}(\omega_2+\omega_1,p_3+\theta^{-1}x_{12}),
\end{equation}
that appears embedded in both Eqs.~(\ref{eq:intcorrection}) and (\ref{eq:div}) will certainly appear at two loops and beyond in diagrams containing the insertion of any marginal operator, unless new divergences beyond one loop are not allowed. Thus, divergent quantum corrections beyond one loop are inherently dangerous in noncommutative theories.

We are led to conjecture that \emph{ the only renormalizable, and
hence, quantum mechanically consistent noncommutative gauge
theories are supersymmetric and do not require renormalization
beyond the one loop order}. It is striking how strong this
statement is. Apparently, noncommutative quantum theories are
extremely sensitive to the UV, and only the softest UV behavior
can be tolerated. In particular, the allowed $2+1$ dimensional
theories include any supersymmetric gauge theory, and the allowed
$3+1$ dimensional theories include both $\mathcal{N}=2$ and
$\mathcal{N}=4$ noncommutative super Yang-Mills (NCSYM) theories
plus some $\mathcal{N}=2$ gauge theories that include matter
hypermultiplets in special representations.

It is interesting that $\mathcal{N}=1$ supersymmetry, or even
$\mathcal{N}=2$ in most cases, does not appear to be enough to
render $3+1$ dimensional noncommutative gauge theories
renormalizable. Moreover, it is tempting to extend our conjecture
to noncommutative theories other than gauge theories. After all,
the intuition that we have gained appears to be quite generic and
most likely applies to any noncommutative theory. For example, it
seems unlikely that noncommutative scalar theory is consistent
quantum mechanically, when nonsupersymmetric gauge theories are not. It even seems reasonable that
renormalizability could fail in $\mathcal{N}=1$ and
$\mathcal{N}=2$ supersymmetric noncommutative theories that do not
include gauge degrees of freedom, in light of the fact that any
extra hypermultiplets of matter that are added to $\mathcal{N}=2$
NCSYM theory must form special representations which do not allow wave function renormalization, so as not to spoil
the theory.

Nonetheless, arguments have been put forward for the renormalizability of many noncommutative
theories that are excluded by our conjecture
\cite{cr:renorm}. Although, these works have involved different
approaches which have encoded UV-IR mixing in one way or another,
none have employed a manifestly dipole construction such as the
matrix formulation. Since the dipole behavior of the elementary
quanta is the fundamental origin of UV-IR mixing in noncommutative
theories, it seems that the physical interpretation that naturally
emerges from the matrix approach is the most reliable.
We believe that this is a tremendous advantage when discussing
renormalizability, because traditionally, the proper treatment of
divergences has resulted from a sound physical interpretation for
their meaning. Of course, the physical content of noncommutative
theories is independent of the language used to discuss them; we
are simply suggesting that the physics is more clear in the matrix representation. In any case,
the difference between our work and \cite{cr:renorm} is the interpretation
and treatment of the dual UV-IR divergences that occur in
noncommutative quantum theories.

Before closing this section, let us make a few comments concerning
the scope of our results from the perspective of string theory.
Our analysis has been limited to noncommutative gauge theories,
and given these dipole degrees of freedom, we have shown that both
a sufficiently high degree of supersymmetry and low spacetime
dimensionality is necessary to ensure the quantum mechanical
consistency of the theory. However, this result does not imply
that other degrees of freedom cannot be added to the theory to
give a consistent UV completion. For example, \cite{rg:ncos} shows
that there is an UV completion of some higher dimensional NCSYM
theories in the form of noncommutative open string theories, in
which the closed string sector has decoupled but there are still
stringy modes from the open string sector that remain. Yet another
distinct possibility is that, in the case of theories with a
lesser degree of supersymmetry, there could be some closed string
modes that survive the decoupling limit and render the theory
consistent \cite{ar:closed}. The point is that any noncommutative
theory that emerges from a decoupling limit of string theory will
be consistent; our results only imply that the decoupled theory
can conceivably be a noncommutative gauge theory in only very
special cases. String theory aside, however, the renormalizability
of noncommutative quantum field theories is interesting in its own
right.

\section{Discussion and outlook}

In this work, we have continued to develop noncommutative gauge
theory in the matrix formulation. After reviewing the dipole
interpretation of the matrix approach, we determined the general
gauge invariant form of perturbative corrections to the quantum
effective action that result from integrating out UV states. We
then studied the divergence structure of these quantum
corrections, which revealed new divergences, having dual UV and IR
interpretations, that are unique to noncommutative theories. These
UV-IR divergences, which appear at two loops and beyond, were
found to represent dangerous quantum corrections that generally
spoil the renormalizability of the theory. Thus, we were led to
conjecture that only supersymmetric noncommutative theories that
do not receive quantum corrections beyond one-loop are renormalizable, and hence, quantum mechanically
consistent.

Furthermore, beyond specific results concerning renormalizability,
it is interesting that the consequences of UV-IR mixing in
noncommutative quantum theories so profoundly affect the long
distance behavior: in $3+1$ dimensions or higher, UV quantum corrections generally introduce new
long distance interactions that grow stronger and stronger, even
if the gauge coupling is small. In the end, the lesson that we
learn is that noncommutative quantum theories exhibit an extreme
sensitivity to the UV, such that only the softest UV behavior can
be tolerated.

However, there still remain a number of open questions concerning
noncommutative quantum field theories. Perhaps, the most glaring
omission in our understanding is the nature of quantum effects due
to IR states in the theory. This important problem represents an
interesting avenue to pursue in future studies. In particular, it
would be very satisfying if both the UV and IR dynamics could be
encoded together in a unified framework. Surely, this would shed
some more light on the fascinating structure of noncommutative
theories.

\begin{acknowledgments}
We thank Li~Jiang, Sonia~Paban, and Willy~Fischler for many
stimulating discussions. This work was supported by NSF grant
PHY-0071512.
\end{acknowledgments}

\appendix

\section{Review of the matrix formulation}\label{ap:review}

For an introduction to the formulation of field theories on
noncommutative spaces, see \cite{ms:nonp,rg:soli}, and for a
complete review of the matrix formulation of noncommutative gauge
theory, see \cite{ns:back,jn:dipo}. Here we will review only the
main results that we will need for the calculations that appear in
this work. Furthermore, we will focus only on the treatment of the
gauge degrees of freedom, the generalization to matter fields,
including fermions, being obvious.

The basic idea of the matrix formulation is that the degrees of
freedom are contained in infinite dimensional time dependent
hermitian matrices $A_0$ and $X^{i}$. In order to separate the
quantum effects from UV and IR states, it will prove convenient to
work in the background field gauge. In this approach, we expand
the fields $A_0=B_0+A$ and $X^{i}=B^{i}+Y^{i}$ where the
background fields, $B_0$ and $B^{i}$, are interpreted as the low
momentum degrees of freedom, while the fluctuating fields, $A$ and
$Y^{i}$, contain the high momentum degrees of freedom. To make the
connection with noncommutative gauge theory, we think of
$B_0=A_0(\hat{x})$ and $B^{i}={\hat{x}}^{i}\otimes {\rlap{1}
\hskip 1.6pt \hbox{1}}_{N\times N}+{\theta}^{ij}A_{j}(\hat{x})$
where $\hat{x}^{i}$ are infinite dimensional time-independent
Hermitian matrices satisfying the algebra of the noncommuting $2p$
plane
\begin{equation}
[\hat{x}^{i},\hat{x}^{j}]=i\theta^{ij}{\rlap{1} \hskip 1.6pt
\hbox{1}}.
\end{equation}

After a suitable choice of gauge fixing, which includes setting
$B_0=0$, the Lagrangian takes the form $L=L_0+L_2+L_3+L_4$ where
{\setlength\arraycolsep{2pt}
\begin{eqnarray}
L_0 & = & {\rm
Tr}\left(\frac{1}{2}\dot{B}^{i2}+\frac{1}{4}[B^{i},B^{j}][B^{i},B^{j}]\right);
\nonumber\\
L_2 & = & {\rm Tr}\left(\frac{1}{2}\dot{Y}^{j2}+\frac{1}{2}[B^{i},Y^{j}]^{2}-\frac{1}{2}{\dot{A}}^{2}-\frac{1}{2}[B^{i},A]^{2}+\dot{\bar{c}}\dot{c} \right. \nonumber\\
 & & \left.{}+[B^{i},\bar{c}][B^{i},c]+[B^{i},B^{j}][Y^{i},Y^{j}]-2i\dot{B}^{i}[A,Y^{i}] \right); \nonumber\\
L_3 & = & {\rm Tr}\left([B^{i},A][A,Y^{i}]+[B^{i},Y^{j}][Y^{i},Y^{j}]+[B^{i},\bar{c}][Y^{i},c] \right. \nonumber\\
 & & \left.{}-i{\dot{Y}}^{i}[A,Y^{i}]-i\dot{\bar{c}}[A,c] \right); \nonumber\\
L_4 & = & {\rm
Tr}\left(\frac{1}{4}[Y^{i},Y^{j}][Y^{i},Y^{j}]-\frac{1}{2}[A,Y^{i}]^{2}\right).
\end{eqnarray}}\\
Notice that we have neglected the linear term in the fluctuating
fields, which generically, is not consistent because we expect that UV quantum
corrections will contribute to the dynamics of the background
field through tadpoles and other effects resulting from the linear
interactions. In the language of perturbation theory, this amounts
to corrections that are both higher order in the gauge coupling
and higher order in derivatives of the background field.
Therefore, in order to consistently ignore the linear term, we must require both that the coupling is
sufficiently weak so that the loop corrections are suppressed and
that the states we integrate out are of sufficiently high momenta
relative to the scale set by the background so that the higher
derivatives are suppressed, as well. We will find that both of
these conditions are met in our approach, so for simplicity sake, we have neglected the linear interactions.

Moving along, from $L_2$, we see that up to commutators and time
derivatives of the background fields, all of the fluctuating
fields, $\Phi=(Y^{i},A,\bar{c},c)$, have similar quadratic terms
of the form
\begin{eqnarray}
& & {\rm
Tr}\left(\frac{1}{2}{\dot{\Phi}}^{2}+\frac{1}{2}[B^{i},\Phi]^{2}\right)\\
& & =\frac{1}{2}\Phi^{T}\left(-{\rlap{1} \hskip 1.6pt
\hbox{1}}\otimes {\rlap{1} \hskip 1.6pt
\hbox{1}}\frac{d^{2}}{dt^{2}}+(B^{i}\otimes {\rlap{1} \hskip 1.6pt
\hbox{1}}-{\rlap{1} \hskip 1.6pt \hbox{1}}\otimes
B^{i})^{2}\right).\nonumber
\end{eqnarray}
For both consistency with the neglect of the linear terms in the
fluctuating fields, as discussed above, and convenience in the
definition of the propagator for the fluctuating fields, we will
treat the commutator and time derivative terms perturbatively
while absorbing the remaining background field dependence into the
definition of the propagator. As discussed in \cite{jn:dipo}, this
choice corresponds to a derivative expansion of the background
field, which makes sense because we will only integrate out high
momentum states.

To lowest order in the derivative expansion, the propagator for
the high momentum modes is
\begin{eqnarray} \label{eq:prop}
\langle\Phi(t)\Phi(t^{\prime})^{T}\rangle &=& \int_{\theta\Lambda}
d^{2p}x
\int_{\Lambda}\frac{d\omega}{2\pi}\int\frac{d^{2p}k}{(2\pi)^{2p}}
e^{-i\omega(t-t^{\prime})}\nonumber\\
& & \quad\times e^{ik\cdot
(x-x^{\prime})}\widetilde{G}(\omega,{\theta}^{-1}(x-x^{\prime}))\nonumber\\
& & \quad\times e^{-ik\cdot B}\otimes e^{ik\cdot B},
\end{eqnarray}
where $\widetilde{G}(\omega,p)=(\omega^{2}-p_iG^{ij}p_j)^{-1}$ is
the momentum space field theory propagator for a massless bosonic state and
$G^{ij}$ is the inverse spatial metric, which we will discuss at
the end of this section. Were we dealing with fermions instead of
bosons, we would simply replace $\widetilde{G}(\omega,p)$ above
with the appropriate propagator for fermi fields.

As discussed in
\cite{jn:dipo}, Eq.~(\ref{eq:prop}) manifestly describes dipole
degrees of freedom whose transverse length is proportional to
their center of mass momentum. In particular, the identification is
that $x$ and $x^{\prime}$ are the endpoints of the dipole in
space, and the center of mass momentum is given by
$p={\theta}^{-1}(x-x^{\prime})$. It is also important to
understand that there is a lower cutoff on the integrals over
$\omega$ and $x$, which reflects the fact that the Wilsonian
integration is only performed over states of high energy and
momentum compared to the background.

In practicality, when we perform calculations, we frequently
encounter traces of the form \cite{mv:mean,jn:dipo}
\begin{eqnarray}\label{eq:rho1}
& & {\rm Tr}\left(\mathcal{O}e^{ik\cdot B}\right)\\
&=& \int d^{2p}x e^{ik\cdot x} {\rm
tr}_N\left(P_{*}e^{i\int_0^{1}d\sigma k\cdot\theta\cdot A(x+\sigma
k\cdot\theta)}*\mathcal{O}(x)\right)\nonumber.
\end{eqnarray}\\
where $\mathcal{O}$ is some operator in the adjoint
representation of the gauge group. We immediately recognize this
object as an open Wilson line with $\mathcal{O}(x)$ attached at
the end. In fact, this structure was essentially guaranteed by the
noncommutative gauge invariance \cite{ni:obse,hl:trek}. Furthermore, we will also frequently
encounter the Fourier transform of the open Wilson lines above,
which we shall denote
\begin{equation}\label{eq:rho2}
\rho_\mathcal{O}(x)=\int\frac{d^{2p}k}{(2\pi)^{2p}}e^{ik\cdot
x}{\rm Tr}\left(\mathcal{O}e^{-ik\cdot B}\right).
\end{equation}\\

Finally, we should mention that we are using a convention of units
in which   $(2\pi)^{2p}det(\theta)=1$ . This convention could
potentially lead to some unfamiliar notions of dimension. What is
more, the situation is further complicated by the fact that the
inverse spatial metric is $G^{ij}=\theta^{ik}\theta^{kj}$.
Nonetheless, the quantities that will appear in our calculations
will all have simple physical interpretations. The rule to follow
is that the magnitude squared of distance $|x_1-x_2|^{2}$ has
dimensions of \emph{energy} squared. The reason is that
\begin{equation}
|x_{12}|^{2}=(\theta^{-1}\cdot x_{12})_i G^{ij}(\theta^{-1}\cdot
x_{12})_j=p_i G^{ij} p_j,
\end{equation}
where we have used the relation between the center of mass
momentum and the position of the end points of the dipole quanta
$p_i=\theta^{-1}_{ij}(x^{j}_1-x^{j}_2)$. Likewise, by the same
argument, the magnitude squared of momenta has dimensions of
length squared.

\section{Leading two-loop calculation}\label{ap:twoloop}

The leading two-loop contribution to the effective action comes
from treating the interaction terms in $L_4$ to first order in
perturbation theory and the interaction terms in $L_3$ to second
order. In \cite{jn:dipo}, the contribution from both the $L_4$
terms as well as the time derivative terms in $L_3$ was computed.
In this section, we will comput the contribution to the effective action from the remaining terms in $L_3$, which can be
represented schematically as
\begin{eqnarray}\label{eq:expect}
& & \langle\int dt_1
\rm{Tr}\left([B,\Phi][\Phi,\Phi](t_1)\right)\nonumber\\
& & \qquad\qquad\times\int dt_2
\rm{Tr}\left([B,\Phi][\Phi,\Phi](t_2)\right)\rangle,
\end{eqnarray}
where $B$ is the low momentum background field and $\Phi$
represents the high momentum fluctuating fields to be integrated
out. The angled brackets denote the vacuum expectation value,
which in our background field language, is the product of the
expectation value of the low momentum ``background fields" and the
expectation value of the high momentum ``fluctuating fields".
Since we are working in the Wilsonian scheme, we evaluate only the
latter quantity and think of the residual expression involving an
expectation value of background fields as originating from an
effective Lagrangian of background fields only.

Furthermore, in the evaluation of Eq.~(\ref{eq:expect}), we neglect the tadpole
contractions, although generally, there will be tadpole
contributions from the linear interaction term in the Lagrangian as well as both the cubic and quartic
interaction terms. However, the linear terms
are suppressed by higher derivatives of the background as we have
already discussed in Appendix~\ref{ap:review}, and in the
supersymmetric theories that  are ultimately of interest to us,
the tadpoles from interaction terms cancel as well. After a
straight forward but somewhat tedious calculation, up to
subleading time derivative and commutator terms we have
\begin{widetext}
\begin{eqnarray}
& & \int dt_1 dt_2 d^{2p}x_1 d^{2p}x_2 d^{2p}x_3 \int
\frac{d\omega_1 d^{2p}k_1}{2\pi(2\pi)^{2p}}\frac{d\omega_2
d^{2p}k_2}{2\pi(2\pi)^{2p}} \frac{d\omega_3
d^{2p}k_3}{2\pi(2\pi)^{2p}} e^{-i\omega_1(t_1-t_2)+ik_1\cdot
x_1}e^{-i\omega_2(t_1-t_2)+ik_2\cdot
x_2}e^{-i\omega_3(t_1-t_2)+ik_3\cdot
x_3}\nonumber\\
& & \qquad\times\widetilde{G}(\omega_1,\theta^{-1}x_1)
\widetilde{G}(\omega_2,\theta^{-1}x_2)
\widetilde{G}(\omega_3,\theta^{-1}x_3)\frac{\partial}{\partial
k_2}\cdot\frac{\partial}{\partial
k_3}\left[\rm{Tr}\left(e^{ik_1\cdot B(t_1)}e^{ik_2\cdot
B(t_1)}\right)\rm{Tr}\left(e^{-ik_1\cdot B(t_1)}e^{ik_3\cdot
B(t_1)}\right)\right.\nonumber\\
& & \qquad\times\left.{}\rm{Tr}\left(e^{-ik_2\cdot
B(t_1)}e^{-ik_3\cdot B(t_1)}\right) -\rm{Tr}\left(e^{ik_1\cdot
B(t_1)}e^{-ik_3\cdot B(t_1)}e^{-ik_2\cdot B(t_1)}e^{-ik_1\cdot
B(t_1)}e^{ik_3\cdot B(t_1)}e^{ik_2\cdot B(t_1)}\right)\right].
\end{eqnarray}
\end{widetext}
Since we only integrate out states with high energy and momentum,
time derivatives and commutators involving the background field
are suppressed. Therefore, to lowest order, we obtain
\begin{eqnarray}
& & \int dt d^{2p}x_1 d^{2p}x_2 d^{2p}x_3 \int\frac{d\omega_1
d^{2p}k_1}{2\pi(2\pi)^{2p}}\frac{d\omega_2
d^{2p}k_2}{2\pi(2\pi)^{2p}} \frac{
d^{2p}k_3}{(2\pi)^{2p}}\nonumber\\
& & \qquad\times e^{ik_1\cdot x_1}e^{ik_2\cdot x_2}e^{ik_3\cdot
x_3}x_2\cdot x_3\nonumber\\
& & \qquad\times\widetilde{G}(\omega_1,\theta^{-1}x_1)
\widetilde{G}(\omega_2,\theta^{-1}x_2)
\widetilde{G}(\omega_1+\omega_2,\theta^{-1}x_3)\nonumber\\
& & \qquad\times\rm{Tr}\left(e^{i(k_1+k_2)\cdot
B(t)}\right)\rm{Tr}\left(e^{-i(k_1-k_3)\cdot
B(t)}\right)\nonumber\\
& & \qquad\times\rm{Tr}\left(e^{-i(k_2+k_3)\cdot B(t)}\right),
\end{eqnarray}
after performing the integral over $\omega_3$ and $t_2$ and
integrating by parts in the $k$ integrals. Note that we have
dropped the single trace term, since as discussed in
\cite{jn:dipo}, it corresponds to a nonplanar matrix diagram,
which is suppressed in the domain of Wilsonian integration.
Finally, upon Fourier transforming back to position space, we are
left with
\begin{eqnarray}
& & \int dt d^{2p}x_1 d^{2p}x_2 d^{2p}x_3
\int\frac{d\omega_1}{2\pi}\frac{d\omega_2}{2\pi}x_{23}\cdot
x_{31}\nonumber\\
& &
\qquad\times\widetilde{G}(\omega_1,\theta^{-1}x_{12})\widetilde{G}(\omega_2,\theta^{-1}x_{23})
\widetilde{G}(\omega_1+\omega_2,\theta^{-1}x_{31})\nonumber\\
& & \qquad\times \rho(x_1,t)\rho(x_2,t)\rho(x_3,t).
\end{eqnarray}

We can now combine the above expression with the result that was
calculated in \cite{jn:dipo} to obtain the leading two-loop
contribution from the second order treatment of all of the cubic
interactions in $L_3$. We get a result proportional to
\begin{eqnarray}\label{eq:totaltwoloop}
& & \int dt d^{2p}x_1 d^{2p}x_2 d^{2p}x_3
\int\frac{d\omega_1}{2\pi}\frac{d\omega_2}{2\pi}(\omega_1\omega_2-x_{12}\cdot
x_{23})\nonumber\\
& & \qquad\times\widetilde{G}(\omega_1,\theta^{-1}x_{12})
\widetilde{G}(\omega_2,\theta^{-1}x_{23})
\widetilde{G}(\omega_1+\omega_2,\theta^{-1}x_{31})\nonumber\\
& & \qquad\times \rho(x_1,t)\rho(x_2,t)\rho(x_3,t).
\end{eqnarray}
Note that we have used the $SO(1,2p)$ symmetry that is present in
the vacuum diagrams in order to determine that the integrand of
the combined expression can only depend on the invariant product
$\omega_1\omega_2-x_{12}\cdot x_{23}$. With this factor present in
the integrand, as in the commutative theories, the two-loop cubic
interactions given by Eq.~(\ref{eq:totaltwoloop}) can be reduced to the
form of Eq.~(\ref{eq:threebody}).

\section{Next to Leading order two loop calculation}\label{ap:ntlotl}

As in the next to leading order one-loop calculation, to get the
precise result we must retain higher order commutators and time
derivatives that were dropped in the derivation of the matrix
propagator (\ref{eq:prop}) as well as the field strength terms in
$L_2$ that were also excluded from the the propagator
\cite{jn:dipo}. However, after developing some intuition in section II for the structure of
terms that can appear, a lengthly calculation is not necessary. We
simply need to keep track of all the distinct possibilities in
which the field strength insertions can appear in the diagrams.
For example, in the case of the quartic diagram, both insertions
can go into one loop or each loop can get a  single insertion
\begin{center}
\begin{minipage}[c]{2.7cm}
\includegraphics[scale=0.4]{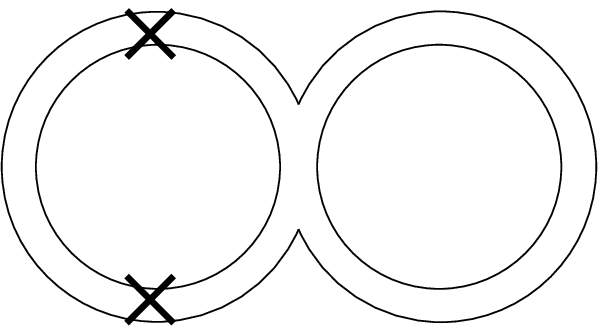}
\end{minipage}%
\begin{minipage}[c]{0.5cm}
or
\end{minipage}%
\begin{minipage}[c]{2.7cm}
\includegraphics[scale=0.4]{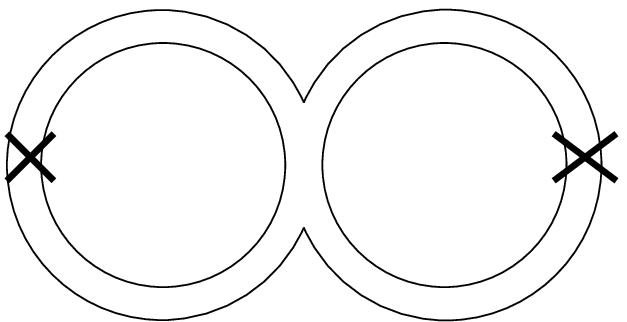}.
\end{minipage}
\end{center}
Of course, each insertion requires an extra field theory
propagator to appear in the corresponding loop, and for each loop,
there are two boundaries in which the insertion can go. Therefore,
the first diagram corresponding to both insertions going into the
same loop contributes two terms
\begin{eqnarray}\label{eq:leading}
& & \int d^{2p}x_1d^{2p}x_2d^{2p}x_3\int
d\omega_1d\omega_2\\
& & \qquad\times\widetilde{G}(\omega_1,\theta^{-1}x_{13})^{3}\widetilde{G}(\omega_2,\theta^{-1}x_{23})\nonumber\\
& &
\qquad\times\left[b_1\rho_{FF}(x_1)\rho(x_2)\rho(x_3)+b_2\rho_F(x_1)\rho(x_2)\rho_F(x_3)\right],\nonumber
\end{eqnarray}
while the second diagram corresponding to one insertion going into
each loop contributes three terms
\begin{eqnarray}\label{eq:beta}
& & \int d^{2p}x_1d^{2p}x_2d^{2p}x_3\int
d\omega_1d\omega_2\\
& & \qquad\times\widetilde{G}(\omega_1,\theta^{-1}x_{13})^{2}\widetilde{G}(\omega_2,\theta^{-1}x_{23})^{2}\nonumber\\
& &
\qquad\times\left[b_3\rho_F(x_1)\rho(x_2)_F\rho(x_3)+b_4\rho(x_1)\rho(x_2)\rho_{FF}(x_3)\right.\nonumber\\
& &
\qquad\qquad\qquad\qquad\qquad\qquad\left.{}+b_5\rho_F(x_1)\rho(x_2)\rho_F(x_3)\right].\nonumber
\end{eqnarray}

There are many more possibilities in the case of the cubic graph.
For simplicity, we will enumerate them in four propagator form,
which is obtained after the cancellation of one propagator by the
momenta in the numerator of the integrand (see
Eq.~(\ref{eq:totaltwoloop})). It is not difficult to see that
there are three combinations of field theory propagators that
emerge: the first is similar to Eq.~(\ref{eq:leading}), the second
is similar to Eq.~(\ref{eq:beta}), and the third  is given by
\begin{eqnarray}\label{eq:dangerous}
& & \int d^{2p}x_1d^{2p}x_2d^{2p}x_3\int d\omega_1d\omega_2\\
& & \qquad\times\widetilde{G}(\omega_1,\theta^{-1}x_{12})^{2}\widetilde{G}(\omega_2,\theta^{-1}x_{23})\widetilde{G}(\omega_1+\omega_2,\theta^{-1}x_{13})\nonumber\\
& &
\qquad\times\left[b_6\rho_{FF}(x_1)\rho(x_2)\rho(x_3)+b_7\rho_F(x_1)\rho_F(x_2)\rho(x_3)\right.\nonumber\\
& &
\qquad\qquad\qquad\qquad\qquad\qquad\left.{}+b_8\rho_F(x_1)\rho(x_2)\rho_F(x_3)\right].\nonumber
\end{eqnarray}

Let us now discuss the meaning of these expressions in the context
of supersymmetric theories. First of all, consider the terms of
the form Eq.~(\ref{eq:leading}). The planar subdiagram
corresponding to the constant term of $\rho(x_2)$, contributes
one-loop divergent mass corrections to the propagators appearing in
Eq.~(\ref{eq:ntlo}). As discussed in section III, in
supersymmetric theories,  contributions of this type will always cancel.

Another obvious group is the combination given by
Eq.~(\ref{eq:beta}). By isolating the constant term of
$\rho(x_2)$, we recover the UV divergent one-loop correction to
the gauge coupling that multiplies Eq.~(\ref{eq:ntlo}). Of course,
these types of corrections are required by renormalizability and
pose no threat to the consistency of the theory.  Moreover, the
constant term of $\rho(x_3)$ gives rise to a benign quantum
correction to the $\rho_F\rho_F$ interaction in
Eq.~(\ref{eq:ntlo}).

Finally, we arrive at the dangerous group of terms given by
Eq.~(\ref{eq:dangerous}). Although, the constant term of
$\rho(x_2)$ gives a harmless quantum correction to the interaction
strength of Eq.~(\ref{eq:ntlo}), the constant term of $\rho(x_3)$
leads to the same type of dangerous UV-IR divergent correction
that was discussed at length in section III. Therefore,
renormalizability demands that these
contributions vanish.


\begin{thebibliography}{99}


\bibitem{sw:stri} N.~Seiberg and E.~Witten, J. High Energy Phys. \textbf{09},
032 (1999).
\bibitem{sb:magn} M.~M.~Sheikh-Jabbari, Phys. Lett. B \textbf{455}, 129 (1999);
D.~Bigatti and L.~Susskind, Phys. Rev. D \textbf{62}, 066004 (2000);
Z.~Yin, Phys. Lett. B \textbf{466}, 234 (1999);
H.~Liu and J.~Michelson, Phys. Rev. D \textbf{62}, 066003 (2000).
\bibitem{ms:nonp} S.~Minwalla, M.~Van Raamsdonk and N.~Seiberg, J. High Energy Phys. \textbf{02}, 020
(2000).
\bibitem{mr:uvdiv} C.P.~Martin and D.~Sanchez-Ruiz, Phys. Rev.
Lett. \textbf{83}, 476 (1999); T.~Krajewski and R.~Wulkenhaar,
Int. J. Mod. Phys. A \textbf{15}, 1011 (2000).
\bibitem{cm:dual} C.P.~Martin and F.R.~Ruiz, Nucl. Phys.
\textbf{B597}, 197 (2001).
\bibitem{cr:renorm} I.~Chepelev and R.~Roiban, J. High Energy
Phys. \textbf{05}, 037 (2000); H.~Grosse, T.~Krajewski, and
R.~Wulkenhaar     ; H.~Girotti, M.~Gomes, V.~Rivelles, and
A.~da~Silva, Nucl. Phys. \textbf{B587}, 299 (2000); I.~Chepelev
and R.~Roiban, J. High Energy Phys. \textbf{03}, 001 (2001);
A.~Bichl, \textit{et al}, J. High Energy Phys. \textbf{06}, 013
(2001); S.~Sarkar, J. High Energy Phys. \textbf{06}, 003 (2002).
\bibitem{mh:irdiv} M.~Hayakawa, Phys. Lett. B \textbf{478}, 394 (2000).
\bibitem{rr:gaugefix} F.R.~Ruiz, Phys. Lett. B \textbf{502}, 274
(2001).
\bibitem{vv:wilson}  V.~V.~Khoze and G.~Travaglini, J. High Energy Phys. \textbf{01}, 026
(2001).
\bibitem{gp:wilsonrg} L.~Griguolo and M.~Pietroni, J. High Energy
Phys. \textbf{05}, 032 (2001).
\bibitem{cw:statmech} G.H.~Chen and Y.S.~Yu, Nucl. Phys.
\textbf{B622}, 189 (2002).
\bibitem{ki:inte} Y.~Kiem, S.~Lee, S.~J.~Rey and H.~T.~Sato, hep-th/0110215;
Y.~Kiem, S.~J.~Rey, H.~T.~Sato and J.~T.~Yee, Phys. Rev. D
\textbf{65}, 026002 (2002); hep-th/0107106; Y.~Kiem, S.~S.~Kim,
S.~J.~Rey and H.~T.~Sato, hep-th/0110066.
\bibitem{ar:uvir} A.~Armoni and E.~Lopez, Nucl. Phys. \textbf{B632}, 240 (2002).
\bibitem{hl:trek} H.~Liu and J.~Michelson, Nucl. Phys. \textbf{B614}, 279 (2001);
H.~Liu, \textit{ibid}. \textbf{B614}, 305 (2001).
\bibitem{mv:mean} M.~Van~Raamsdonk, J. High Energy Phys. \textbf{11}, 006 (2001).
\bibitem{jn:dipo} L.~Jiang and E.~Nicholson, Phys. Rev. D
\textbf{65}, 105020 (2002).
\bibitem{kk:bilocal} S.~Iso, H.~Kawai, and Y.~Kitazawa, Nucl.
Phys. \textbf{B576}, 375 (2000).
\bibitem{ni:obse} N.~Ishibashi, S.~Iso, H.~Kawai and Y.~Kitazawa, Nucl. Phys. \textbf{B573}, 573 (2000);
S.~J.~Rey and R.~von~Unge, Phys. Lett. B \textbf{499}, 215 (2001);
S.~Das and S.~J.~Rey, Nucl. Phys. \textbf{B590}, 453 (2000);
D.~Gross, A.~Hashimoto and N.~Itzhaki, Adv. Theor. Math. Phys.
\textbf{4}, 893 (2000).
\bibitem{ns:back} N.~Seiberg, J. High Energy Phys. \textbf{09}, 003 (2000).
\bibitem{ls:gaug} A.~Matusis, L.~Susskind and N.~Toumbas, J. High Energy Phys.
\textbf{12}, 002 (2000).
\bibitem{dz:susy} D.~Zanon, Phys. Lett. B \textbf{502}, 265
(2001).
\bibitem{rg:ncos} R.~Gopakumar, S.~Minwalla, N.~Seiberg, and
A.~Strominger, J. High Energy Phys. \textbf{08}, 008 (2000).
\bibitem{ar:closed} A.~Rajaraman and M.~Rozali, J. High Energy
Phys. \textbf{04}, 033 (2000).
\bibitem{rg:soli} R.~Gopakumar, S.~Minwalla and A.~Strominger, J. High Energy Phys. \textbf{05}, 020 (2000).


\end{thebibliography}
\end{document}